\documentclass[11pt]{article}
\usepackage{graphicx}
\usepackage{amsmath,amssymb}
\usepackage{color}
\usepackage{tikz}

\def\1{{\bf 1}}

\def\be{\begin{equation}}
\def\ee{\end{equation}}

\begin{document}
\noindent {\Large \bf The polar clasps of a bank vole PrP(168--176) prion protofibril revisiting}\\

\bigskip

\noindent {\Large Jiapu Zhang}\\
{\small {\it
\noindent Centre of Informatics and Applied Optimisation, The Federation University Australia, Mount Helen Campus, Mount Helen, Ballarat, Victoria 3353, Australia;

\noindent $^\text{*}$Tel: +61-3-5327 6335; Email: j.zhang@federation.edu.au\\
}}

\noindent {\bf Abstract:} 
On 2018-01-17 two electron crystallography structures (with PDB entries 6AXZ, 6BTK) on a prion protofibril of bank vole PrP(168-176) (a segment in the PrP $\beta$2-$\alpha$2 loop) were released into the PDB Bank. The paper published by [Nat Struct Mol Biol 25(2):131-134 (2018)] reports some polar clasps for these two crystal structures, and "an intersheet hydrogen bond between Tyr169 and the backbone carbonyl of Asn171 on an opposing strand." - this hydrogen bond is not between the neighbouring Chain B and Chain A directly. In addition, by revisiting the polar clasps, we found another two hydrogen bonds (B.Asn171@H-A.Gln172@OE1, B.Tyr169@OH-A.Gln172@N) between the strand A of one sheet and the opposing strand B of the mating sheet. For the neighbouring two single $\beta$-sheets AB, the two new hydrogen bonds are completely different from the experimental one (an intersheet hydrogen bond between Tyr169 and the backbone carbonyl of Asn171 on an opposing strand) in [Nat Struct Mol Biol 25(2):131-134 (2018)].


\section{Introduction}
Prion diseases were caused by the conformational change from normal cellular prion protein PrP$^C$ into diseased infectious prion PrP$^{Sc}$. Numerous structures of PrP$^C$ are known and have been deposited into the Protein Data Bank (rcsb.org) (PDB). However, the atomic structure of the infectious, protease-resistant, $\beta$-sheet-rich and fibrillar mammalian PrP$^{Sc}$ remains unknown, being due to the unstable, noncrystalline and insoluble nature of the fibrils. Only during recent a decade the lab of Eisenberg DS at University of California Los Angeles and other labs are producing the atomic structures of some segments of PrP$^{Sc}$. On 2018-01-17 two electron crystallography structures (with PDB entries 6AXZ, 6BTK) on a prion protofibril of bank vole PrP(168--176) (a segment in the PrP  $\beta$2-$\alpha$2 loop) were released into the PDB Bank. The structures belong to the ordinary Class 2 steric zipper with $\beta$-strands parallel and sheets pair front to back \cite{sawaya_etal2007}, but they have an outstanding hallmark: within a sheet there are very rich hydrogen bonds (HBs) - even constructing several `circles' by polar residues Gln, Asn and Tyr of the two neighboring $\beta$-strands of a sheet; and between two neighboring $\beta$-sheets (if each is with one and only one $\beta$-strand) there are two additional HBs (B.Asn171@H--A.Gln172@OE1, B.Tyr169@OH--A.Gln172@N) too -  besides the one of \cite{gallagher_etal2018}. The authors named this character as `polar clasps'. This brief paper is revisiting the `polar clasps' from the points of view in crystal mathematical formulas and mathematical optimization. Detailed HBs will be presented and illuminated for the `polar clasps'.

First we introduce the definition of `polar clasps' in \cite{gallagher_etal2018}. `Polar clasp' is a motif whose neighboring polar ladders in the PrP$^{Sc}$ are linked by HBs within a strand. Stacks of aromatic residues Phe175 and Tyr169 will shield these clasps at the core of PrP$^{Sc}$.

In \cite{zhang2016}, the author showed us in Year 2016 a table listing all the PrP segments with cross-$\beta$ structures and each model is with its mathematical formulas and a color photo. Now we insert two elements ({\it (21)--(22)}) of bank vole (Myodes glareolus) PrP(168--176) of Year 2018 into the table and produce a new table, Tab. \ref{cross_beta_PrP_segments_structures}. We organize this paper as follows. Firstly, we present the approaches and tools that will be used in this paper and our major discovery about (molecular modelling) optimization skills will be given in the {\it Methods Section}. In the {\it Results and Discussion Section}, for models {\it (21)--(22)}, {\it the first Subsection} will present mathematical formulas accompanied by graphs, then {\it the second Subsection} will revisit HBs of the `polar clasps'. Lastly, the {\it Summary Section} gives four concluding remarks.
\begin{table}[h!]
\caption{\textsf{The cross-$\beta$ structures known in the PDB Bank of PrP segments:}}
\centering
{\small
\begin{tabular}{l                           |l                    |c                    |c} \hline
                           PrP segment      &Species              &PDB ID               &Class of the cross-$\beta$\\ \hline \hline
                {\it (01)} PrP(126--131)    &human                &4TUT                 &Class 7$^{\cite{sawaya_etal2007,yu_etal2015}}$\\ 
                {\it (02)}                  &human-V129           &4UBY                 &Class 8$^{\cite{sawaya_etal2007,yu_etal2015}}$\\
                {\it (03)}                  &human-L129           &4UBZ                 &Class 8$^{\cite{sawaya_etal2007,yu_etal2015}}$\\ \hline
                {\it (04)} PrP(126--132)    &human-L129           &4W5L                 &Class 8$^{\cite{sawaya_etal2007,yu_etal2015}}$\\
                {\it (05)}                  &human                &4W5M                 &Class 8$^{\cite{sawaya_etal2007,yu_etal2015}}$\\
                {\it (06)}                  &human-V129           &4W5P                 &Class 8$^{\cite{sawaya_etal2007,yu_etal2015}}$\\ \hline
                {\it (07)} PrP(127--132)    &human                &4WBU                 &Class 8$^{\cite{sawaya_etal2007,yu_etal2015}}$\\
                {\it (08)}                  &human-V129           &4WBV                 &Class 8$^{\cite{sawaya_etal2007,yu_etal2015}}$\\
                {\it (09)}                  &human                &3MD4                 &antiparallel (P 2$_1$ 2$_1$ 2$_1$)\\
                {\it (10)}                  &human-V129           &3MD5                 &parallel (P 1 2$_1$ 1)\\
                {\it (11)}                  &human                &3NHC                 &Class 8$^{\cite{sawaya_etal2007}}$\\
                {\it (12)}                  &human-V129           &3NHD                 &Class 8$^{\cite{sawaya_etal2007}}$\\ \hline
                {\it (13)} PrP(127--133)    &human                &4W5Y                 &Class 6$^{\cite{sawaya_etal2007,yu_etal2015}}$\\
                {\it (14)}                  &human-V129           &4W67                 &Class 6$^{\cite{sawaya_etal2007,yu_etal2015}}$\\
                {\it (15)}                  &human-L129           &4W71                 &Class 6$^{\cite{sawaya_etal2007,yu_etal2015}}$\\ \hline
                {\it (16)} PrP(137--142)    &mouse                &3NVG                 &Class 1$^{\cite{sawaya_etal2007}}$\\ \hline
                {\it (17)} PrP(137--143)    &mouse                &3NVH                 &Class 1$^{\cite{sawaya_etal2007}}$\\ \hline
                {\it (18)} PrP(138--143)    &Syrian hamster       &3NVE                 &Class 6$^{\cite{sawaya_etal2007}}$\\
                {\it (19)}                  &human                &3NVF                 &Class 1$^{\cite{sawaya_etal2007}}$\\ \hline
                {\it (20)} PrP(142--166)    &sheep                &1G04                 &$\beta$-hairpin$^{\cite{kozin_etal2001}}$\\ \hline
                {\it (21)} PrP(168--176)    &bank vole            &6AXZ                 &Class 2$^{\cite{gallagher_etal2018, sawaya_etal2007}}$\\
                {\it (22)}                  &bank vole            &6BTK                 &Class 2$^{\cite{gallagher_etal2018, sawaya_etal2007}}$\\ \hline
                {\it (23)} PrP(170--175)    &human                &2OL9                 &Class 2$^{\cite{sawaya_etal2007}}$\\
                {\it (24)}                  &elk                  &3FVA                 &Class 1$^{\cite{sawaya_etal2007}}$\\ \hline
{\it (25)} PrP(177--182, 211--216) &human   &4E1I &$\beta$-prism I fold$^{\cite{apostol_etal2013}}$ (P 2$_1$ 2$_1$ 2$_1$)\\
{\it (26)}                         &human   &4E1H &$\beta$-prism I fold$^{\cite{apostol_etal2013}}$ (P 2$_1$ 2$_1$ 2$_1$)\\ \hline
\end{tabular} 
} \label{cross_beta_PrP_segments_structures}
\end{table}

\section{Methods - The Molecular Modelling Approaches and Tools}
The molecular modelling approaches used in the paper are the deep analyses for the files from the PDB library, optimization theory and the hybrid of optimization methods. The tools used are JSmol (http://www.jmol.org) - an open-source JavaScript computer software for molecular modelling chemical structures in 3-dimensions, VMD \cite{humphrey_etal1996} (where the VMD is the most recent version VMD1.9.3Win32cuda downloaded from http://www.ks.uiuc.edu/Research/vmd/) - a molecular visualization program for displaying, animating, and analysing large biomolecular systems using 3-D graphics and built-in scripting, and the Swiss-PdbViewer 4.1.0 (https://spdbv.vital-it.ch/) \cite{guex_etal2009, guex_etal1997} - a free package for comparative protein modelling.

Optimization plays important roles in molecular modelling, and molecular simulations of molecular dynamics and/or quantum mechanics/molecular mechanics. Large-scale optimization computations have drawn considerable attention. The existing local/global optimization techniques effectively solve many problems when the number of variables is not very large and, as a rule, fail to solve many large-scale problems. The study of new algorithms which allow one to solve large-scale optimization problem is very important. One technique is to use hybrid of local/global and local/global search algorithms. This paper advocates the use of hybrid algorithmic approaches \cite{zhang2011}. It is clearly possible that different optimization strategies would have made different results. In the second subsection of the Results and Discussion section we will use the hybrid of local search optimization methods Steepest Descent (SD) method + Conjugate Gradient (CG) method + Steepest Descent (SD) method when using Swiss-PdbViewer 4.1.0.

\section{Results and discussion}
\subsection{Mathematical formulas and graphs}
In this subsection, we will give the mathematical formulas for the fibrils of bank vole PrP(168--176) with PDB entries 6AXZ and 6BTK respectively, then we will illuminate their respective photos. The mathematical formulas can be gotten after reading the PDB files at websites:\\
\centerline{https://files.rcsb.org/view/6AXZ.pdb, https://files.rcsb.org/view/6BTK.pdb,}
where the coordinates of the A Chain were given and we were told clearly ``APPLY THE FOLLOWING TO CHAINS: A" to get other Chains. Their photos can be gotten from the websites https://www.rcsb.org/3d-view/6AXZ/1, https://www.rcsb.org/3d-view/6BTK/1 respectively by selecting the viewer JSmol (JavaScript).
\subsubsection{PDB entry 6AXZ -- bank vole PrP(168--176)}
\begin{figure}[h!]
\centerline{
\includegraphics[width=5.2in]{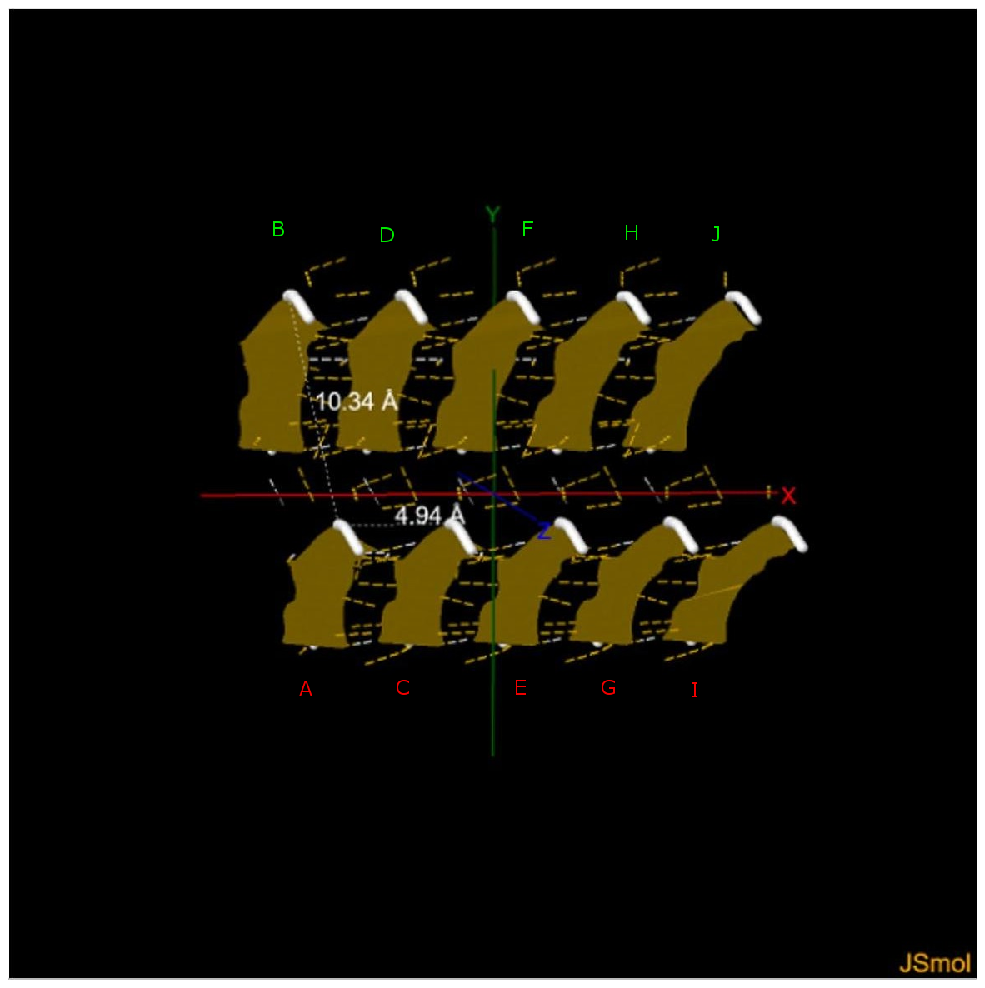}
}
\caption{\textsf{ Protein fibril structure of QYNNQNNFV segment 168--176 from bank vole prion (PDB entry 6AXZ). The  dashed lines denote the HBs. A, B, ..., I, J denote the 10 chains of the fibril.}} \label{revisiting_6AXZ}
\end{figure}

Model {\it (21)}: 
In Fig. \ref{revisiting_6AXZ} we see that B chain (i.e. $\beta$-sheet 2) of 6AXZ.pdb can be obtained from A chain (i.e. $\beta$-sheet 1) by
\begin{equation}
B = A + \left( \begin{array}{c}
-2.18510\\
10.10648\\ 
0\end{array} \right), \label{21_6AXZ_ab}
\end{equation}
and other chains can be got by
\begin{equation}
C (D)= A (B)+ \left( \begin{array}{c}
4.94\\
0\\ 
0\end{array} \right), \label{21_6AXZ_cd}
\end{equation}
\begin{equation}
E (F) = A (B) +\left( \begin{array}{c}
2*4.94\\
0\\ 
0\end{array} \right), \label{21_6AXZ_ef}
\end{equation}
\begin{equation}
G (H) = A (B) +\left( \begin{array}{c} 
3*4.94\\
0\\ 
0\end{array} \right), \label{21_6AXZ_gh}
\end{equation}
\begin{equation}
I (J) = A (B) + \left( \begin{array}{c} 
4*4.94\\
0\\ 
0\end{array} \right). \label{21_6AXZ_ij}
\end{equation}

\subsubsection{PDB entry 6BTK -- bank vole PrP(168--176)}
\begin{figure}[h!]
\centerline{
\includegraphics[width=5.2in]{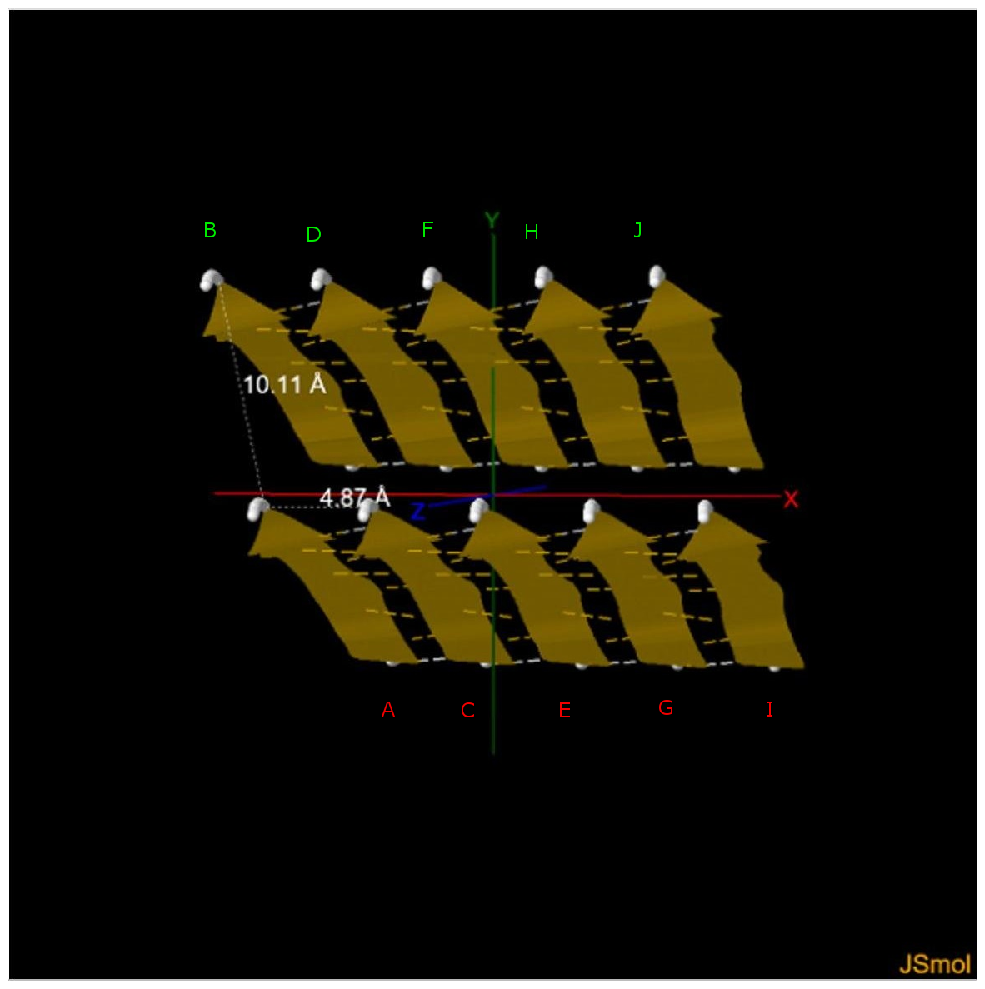}
}
\caption{\textsf{ Protein fibril structure of QYNNQNNFV segment 168--176 from bank vole prion (PDB entry 6BTK). The  dashed lines denote the HBs. A, B, ..., I, J denote the 10 chains of the fibril.}} \label{revisiting_6BTK}
\end{figure}

Model {\it (22)}: 
In Fig. \ref{revisiting_6BTK} we see that B chain (i.e. $\beta$-sheet 2) of 6BTK.pdb can be obtained from A chain (i.e. $\beta$-sheet 1) by
\begin{equation}
B = A + \left( \begin{array}{c}
-2.04266\\
 9.89843\\ 
0\end{array} \right), \label{22_6BTK_ab}
\end{equation}
and other chains can be got by
\begin{equation}
C (D)= A (B)+ \left( \begin{array}{c}
4.874\\
0\\ 
0\end{array} \right), \label{22_6BTK_cd}
\end{equation}
\begin{equation}
E (F) = A (B) +\left( \begin{array}{c}
2*4.874\\
0\\ 
0\end{array} \right), \label{22_6BTK_ef}
\end{equation}
\begin{equation}
G (H) = A (B) +\left( \begin{array}{c} 
3*4.874\\
0\\ 
0\end{array} \right), \label{22_6BTK_gh}
\end{equation}
\begin{equation}
I (J) = A (B) + \left( \begin{array}{c} 
4*4.874\\
0\\ 
0\end{array} \right). \label{22_6BTK_ij}
\end{equation}

\subsection{Revisiting the HBs}
We used the above mathematical formulas and the Figs. \ref{revisiting_6AXZ}-\ref{revisiting_6BTK} to get {\it (I) AC-Chains} and {\it (II) AB-Chains} for 6AXZ.pdb and 6BTK.pdb respectively, and {\it (III) ABCD-Chains} for 6AXZ.pdb. We used VMD to visualize the HBs of {\it (I)--(III)} for each pdb but we cannot find HBs yet (where the HB parameters are the default ones of VMD: Donor-Acceptor Distance Cutoff 3.0 $\mathring{A}$, Angle cutoff 20 degrees). Thus we used Swiss-PdbViewer 4.1.0 (the file SPDBV\_4.10\_PC.zip downloaded from https://spdbv.vital-it.ch/) to optimize {\it (I)--(III)} (Tab. \ref{Optimization}), where we found the hydrogen atoms were added after the optimization by the Swiss-PdbViewer. 

The Swiss-PdbViewer is the most recent version Swiss-PdbViewer 4.1.0 and we set the Energy Minimization Preferences as follows: 100 Steps of Steepest Descent (SD) method, then 100 Steps of Conjugate Gradient (CG) method, and at last 50 Steps of SD method to slightly optimize the model - i.e. here we used the hybrids of SD-CG-SD optimization methods. The optimization considered the Bonds, Angles, Torsions, Improper, Non-bonded, Electrostatic, and we selected the Cutoff as 12.000 $\mathring{A}$. The stopping criteria when delta E between two steps is below 0.050 KJ/mol and when Force acting on any atom is below 10.000 are selected. The computations were done in vacuo with the GROMOS96 43B parameters set, without reaction field, and the energy computations were done with the GROMOS96 implementation of Swiss-PdbViewer 4.1.0.
\begin{table}[h!]
\caption{\textsf{The energy decreases during the optimization:}}
\centering
{\small
\begin{tabular}{l        |l                    |c      |c            |l                        } \hline
         PrP segment     &Species              &PDB ID &Chains       &Energy (KJ/mol) decreases\\ \hline \hline
{\it (21)} PrP(168--176) &bank vole            &6AXZ  &AC            &-2250.996 $\to$ -2264.804 $\to$ -2268.384\\
                         &                     &      &ACE           &-3617.681 $\to$ -3635.713 $\to$ -3646.515\\
                         &                     &      &ABCD          &-4706.212 $\to$ -4715.012 $\to$ -4726.107\\
                         &                     &      &AB            &-1885.567 $\to$ -1901.603 $\to$ -1904.400\\ \hline
{\it (22)}               &bank vole            &6BTK  &AC            &-2268.958 $\to$ -2269.830 $\to$ -2274.719\\
                         &                     &      &ACE           &-3664.371 $\to$ -3674.300 $\to$ -3678.038\\
                         &                     &      &AB            &-1916.021 $\to$ -1919.213 $\to$ -1922.448\\ \hline  
\end{tabular} 
} \label{Optimization}
\end{table}

After the optimization, we got the HBs network for {\it (I)--(III)} respectively (Tabs. \ref{HBs_ACchains}-\ref{HBs_ABCDchains}), where the HB detection threshold parameters are the default ones of Swiss-PdbViewer 4.1.0: when Hydrogens are present,     Min cutoff distance 1.200 $\mathring{A}$, Max cutoff distance 2.760 + 0.050 $\mathring{A}$, Min Cutoff Angle 120.000 degrees; and when Hydrogens are not present, Min cutoff distance 2.195 $\mathring{A}$, Max cutoff distance 3.300 + 0.050 $\mathring{A}$, Min Cutoff Angle 90.000 degrees. 

\begin{figure}[h!]
\centerline{
\includegraphics[width=2.8in]{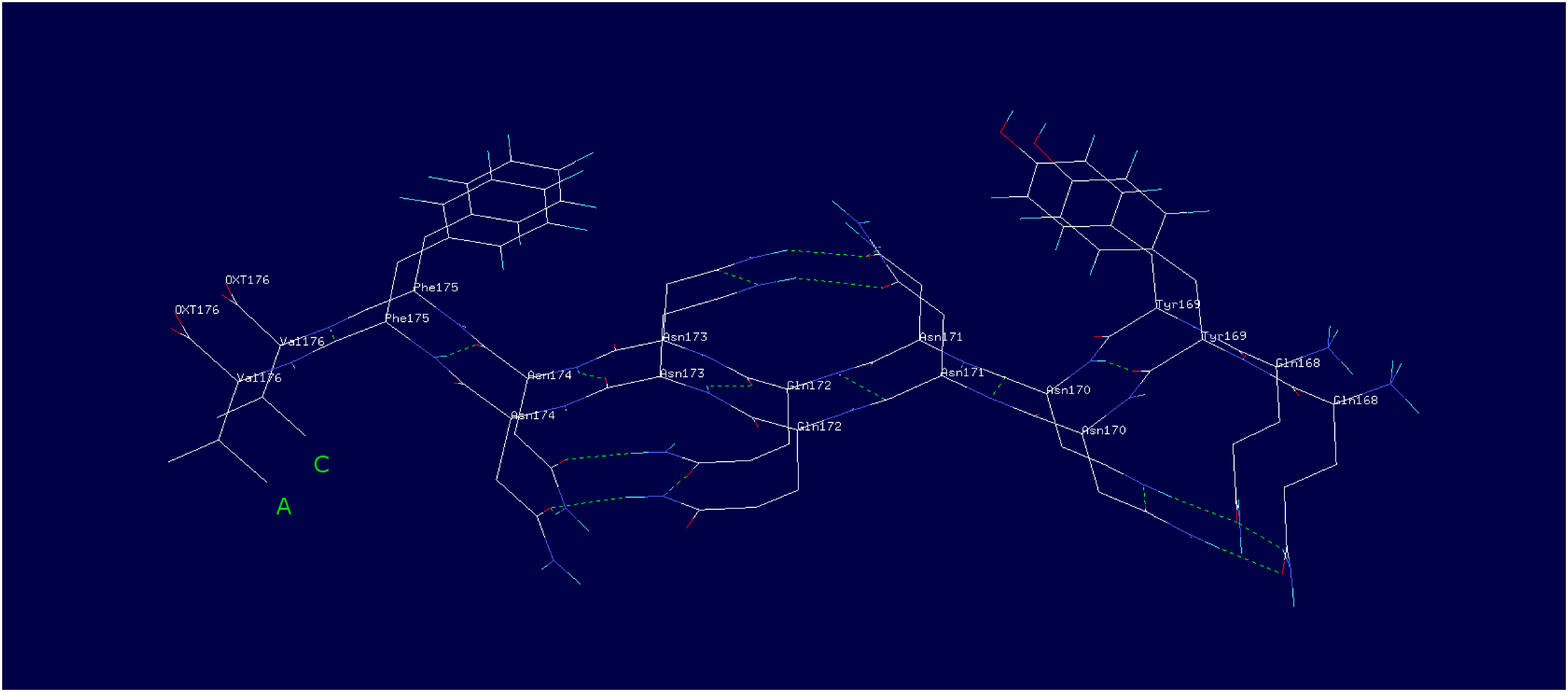}
\includegraphics[width=2.8in]{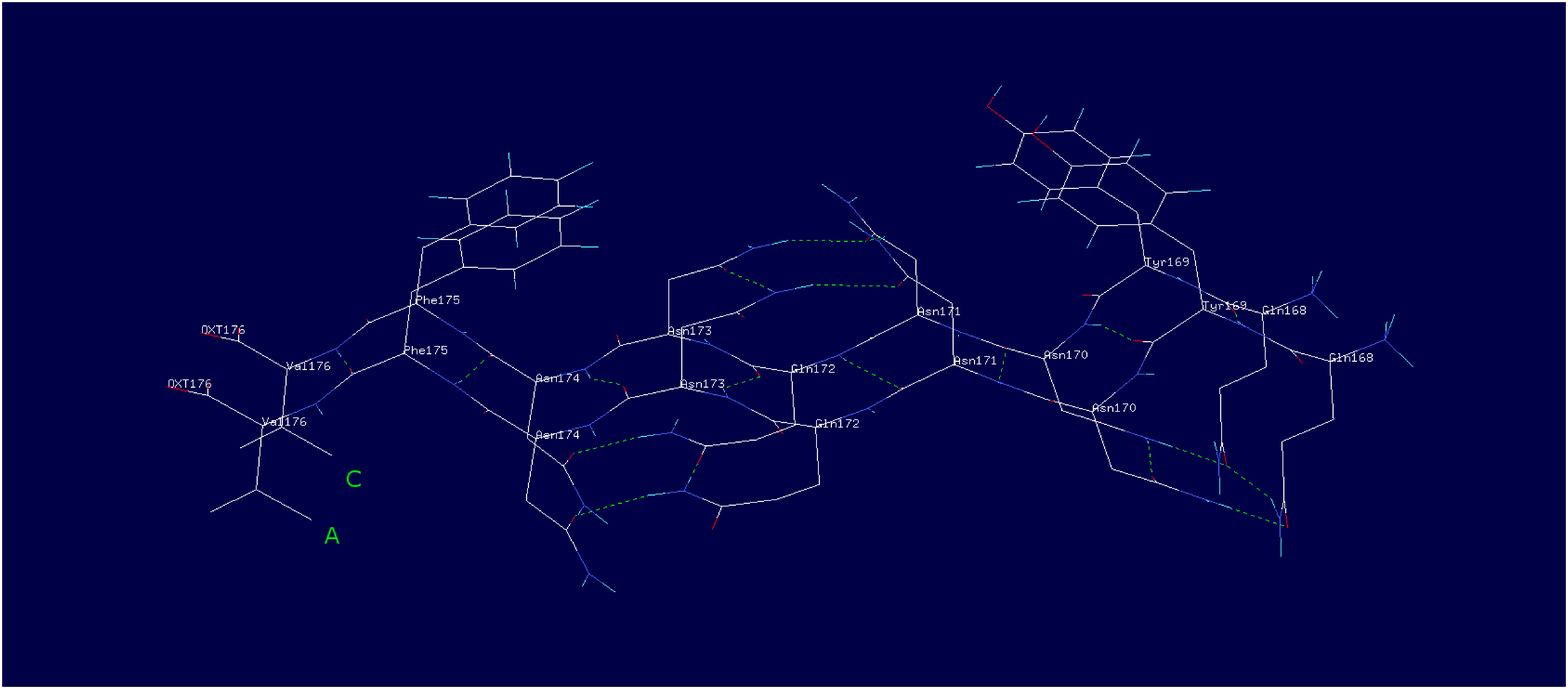}
}
\caption{\textsf{ The optimized AC Chains in the fibril structures of QYNNQNNFV segment 168--176 from bank vole prion (left: 6AXZ, right: 6BTK). The dashed lines denote the HBs. The photos were produced by Swiss-PdbViewer 4.1.0. The green coloured A and C denote the A Chain and C Chain respectively.}} \label{HBs_ACchains_illuminations}
\end{figure}

{\it (I) AC-Chains}. There are 20 HBs for the optimized AC Chains of 6AXZ and 6BTK respectively (Tab. \ref{HBs_ACchains} and Fig. \ref{HBs_ACchains_illuminations}). Here we found 14 interchain HBs, including the 6 interchain HBs in Fig. 2 of \cite{gallagher_etal2018} labeled by blue arrows.  
\begin{table}[h!]
\caption{\textsf{The 20 HBs between the optimized AC Chains or within its each Chain for each pdb model:}}
\centering
{\small
\begin{tabular}{l               |l                          |l}\hline
between C--A Chains             &in A Chain                 &in C Chain\\ \hline \hline
Val176@H--Phe175@O              &Asn174@OD1--Gln172@H       &Asn174@OD1--Gln172@H\\ 
Asn174@O--Phe175@H              &ASN171@OD1--Asn173@H       &Asn171@OD1--Asn173@H\\
Asn174@H--Asn173@O              &Asn170@H--Gln168@OE1       &Asn170@H--Gln168@OE1\\
Asn174@H--Asn174@OD1            &                           &\\
Asn173@OD1--Asn173@H            &                           &\\
Gln172@H--Asn171@O              &                           &\\
Gln172@OE1--Gln172@H            &                           &\\ 
Gln172@O--Asn173@H              &                           &\\
Asn171@OD1--Asn171@H            &                           &\\
Asn170@O--Asn171@H              &                           &\\
Asn170@H--Asn170@OD1            &                           &\\
Asn170@H--Tyr169@O              &                           &\\
Gln168@O--Tyr169@H              &                           &\\
Gln168@OE1--Gln168@H            &                           &\\ \hline  
\end{tabular} 
} \label{HBs_ACchains}
\end{table}

\begin{figure}[h!]
\centerline{
\includegraphics[width=2.8in]{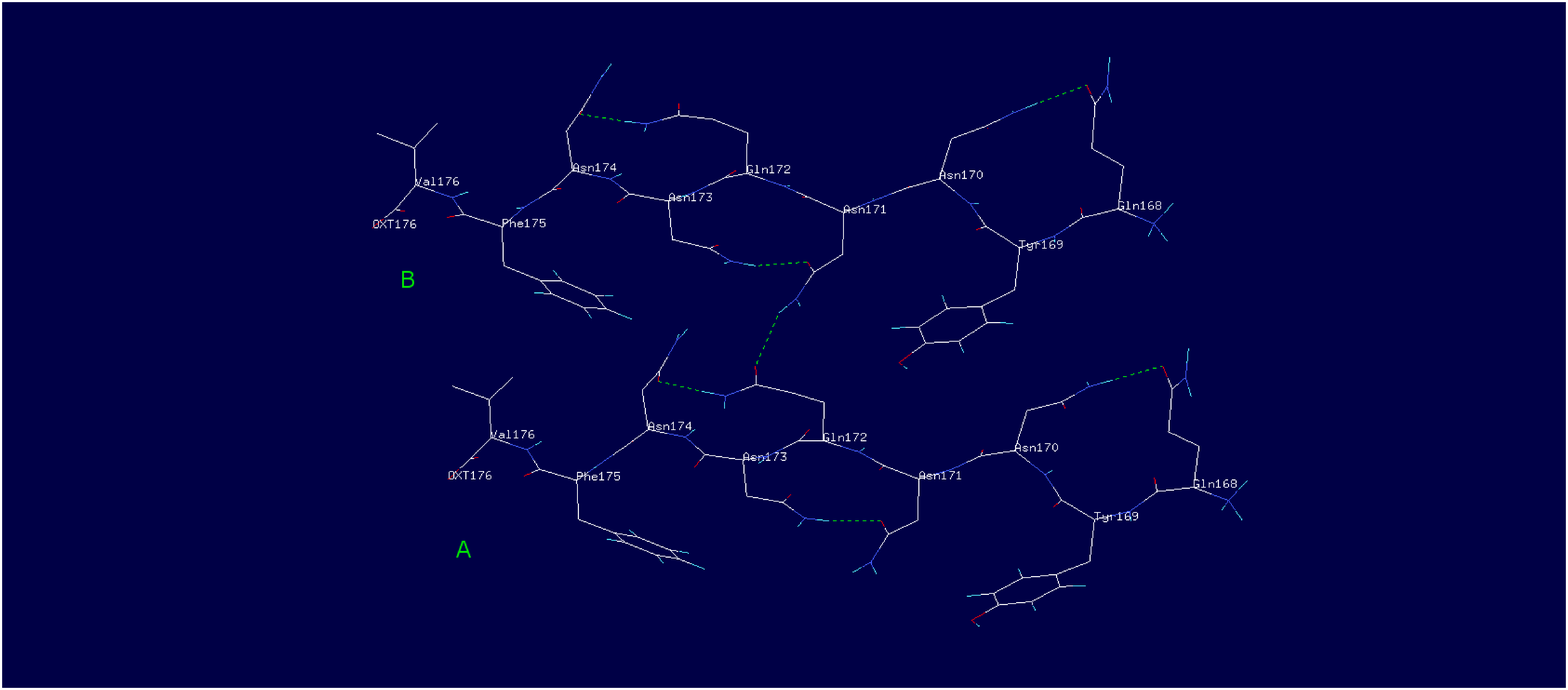}
\includegraphics[width=2.8in]{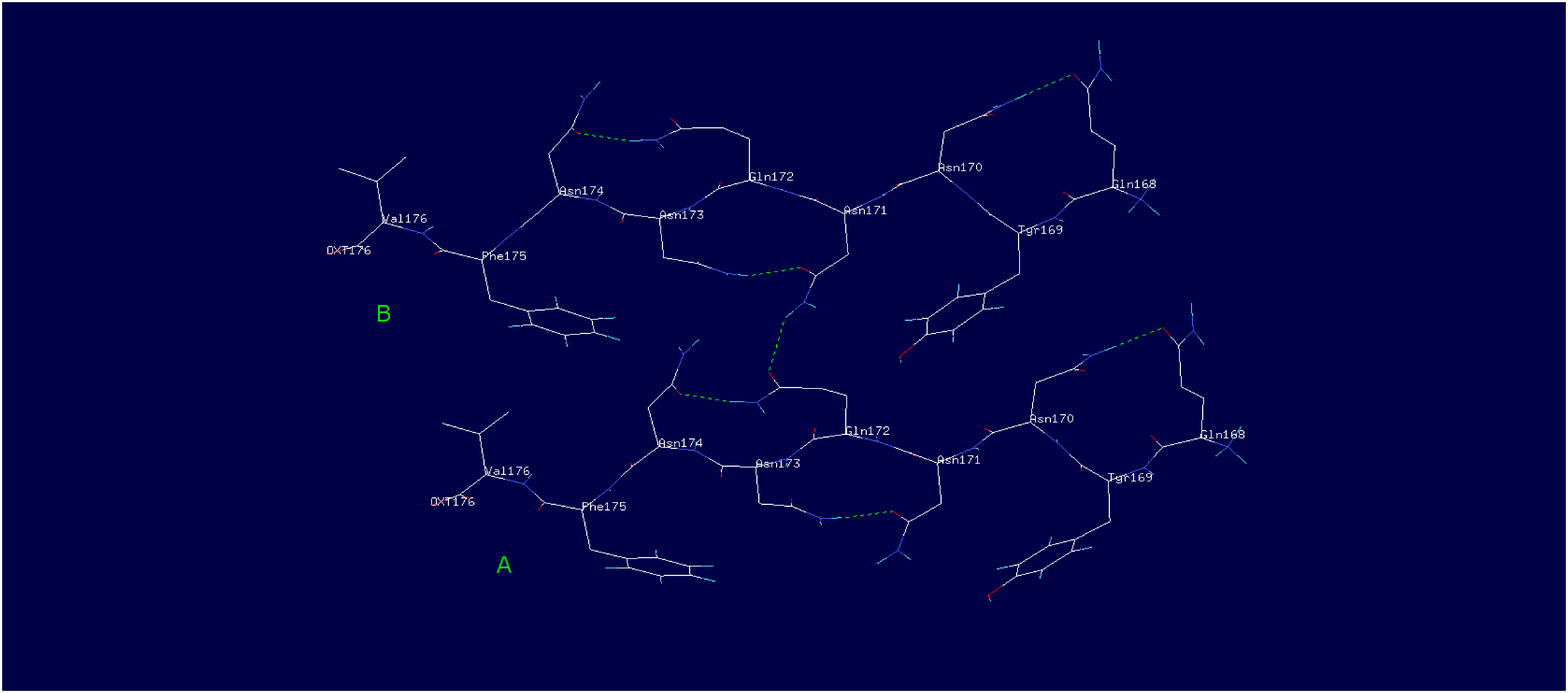}
}
\caption{\textsf{ The optimized AB Chains in the fibril structures of QYNNQNNFV segment 168--176 from bank vole prion (left: 6AXZ, right: 6BTK). The dashed lines denote the HBs. The photos were produced by Swiss-PdbViewer 4.1.0. The green coloured A and B denote the A Chain and B Chain respectively.}} \label{HBs_ABchains_illuminations}
\end{figure}

\begin{figure}[h!]
\centerline{
\includegraphics[width=5.6in]{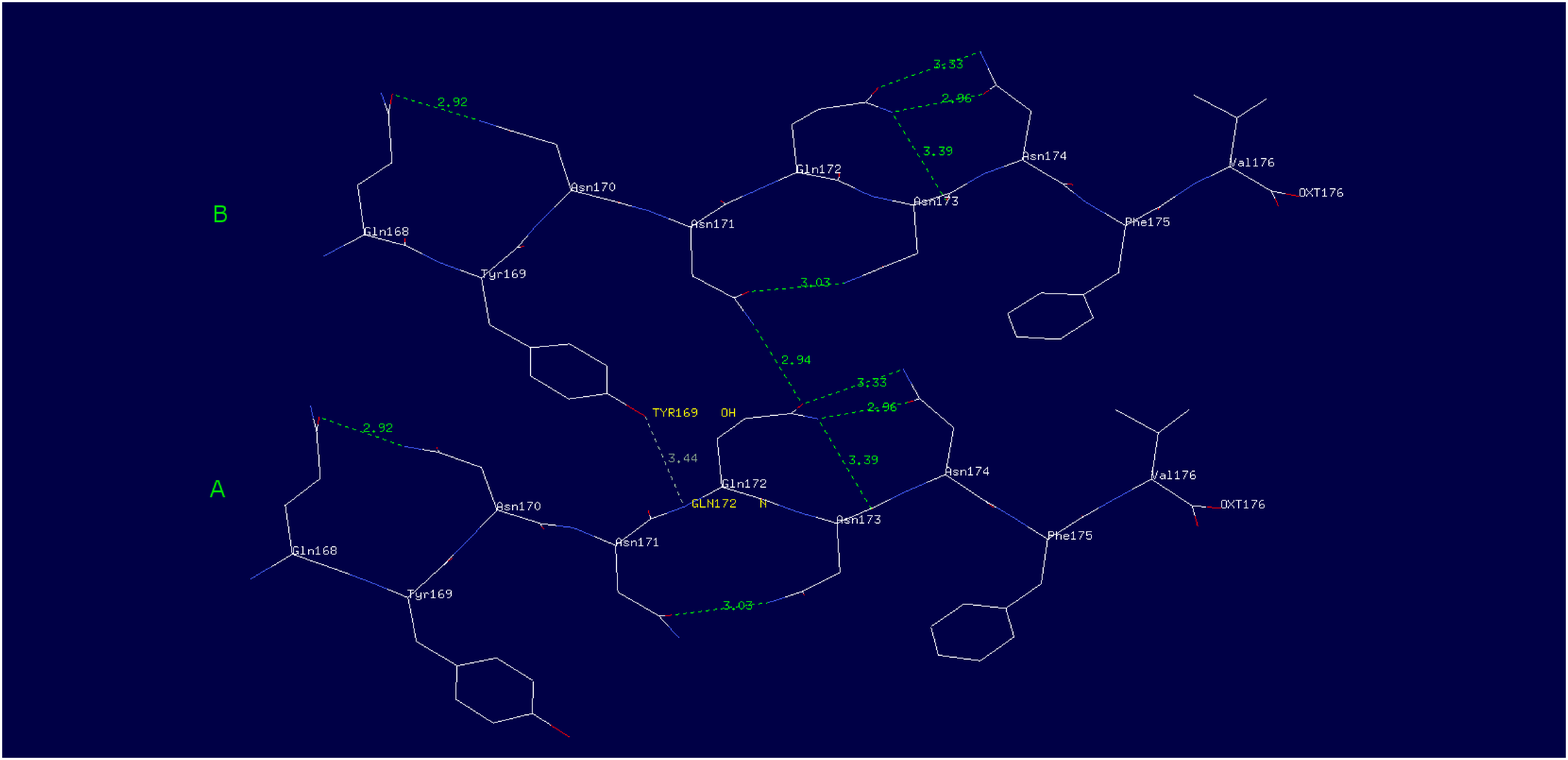}
}
\caption{\textsf{ The HB B.Tyr169@OH--A.Gln172@N between AB Chains (not optimized yet, directly got by mathematical formula (\ref{22_6BTK_ab})) in the fibril structures of QYNNQNNFV segment 168--176 from bank vole prion (6BTK.pdb). The dashed lines denote the HBs. The photo was produced by Swiss-PdbViewer 4.1.0.}} \label{HB_B_Tyr169-OH-A_Gln172-N}
\end{figure}

\begin{figure}[h!]
\centerline{
\includegraphics[width=5.6in]{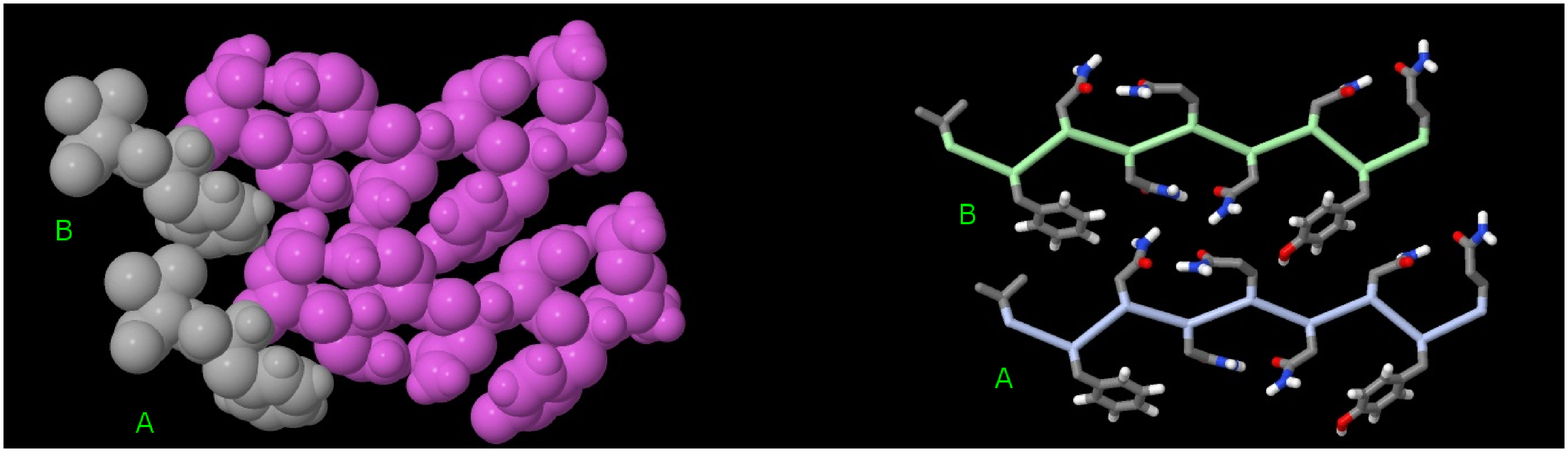}
}
\caption{\textsf{ The distributions of residues of AB Chains in the fibril structures of QYNNQNNFV segment 168--176 from bank vole prion (PDB entry 6AXZ). Pink: polar residues Gln168--Asn174, gray: hydrophobic residues Phe175--Val176.}} \label{6AXZ_ABchains_optimized_Hydrophobic_Polar}
\end{figure}

{\it (II) AB-Chains}. There are 7 HBs for the optimized AB Chains of 6AXZ and 6BTK respectively (Tab. \ref{HBs_ABchains} and Fig. \ref{HBs_ABchains_illuminations}). Here the single intersheet HB Asn171@H--Gln172@OE1 is different from the one in Fig. 2 of \cite{gallagher_etal2018} labeled by a red arrow (a note: {\it if we set the Max distance 3.4 $\mathring{A}$ when Hydrogens are not present, we can find a HB B.Tyr169@OH--A.Gln172@N with distance 3.44 $\mathring{A}$ (Fig. \ref{HB_B_Tyr169-OH-A_Gln172-N}) where the coordinate of B.Tyr169@OH is (-3.491, 1.826, 6.481), the coordinate of A.Gln172@N is (-1.581, -0.424, 8.254), and B Chain is got from A Chain by mathematical formula (\ref{22_6BTK_ab})}. We should notice here the AB Chains are not optimized yet). The intrachain three HBs labeled by orange arrows in Fig. 2 of \cite{gallagher_etal2018} are found here (see Tab. \ref{HBs_ABchains} and Fig. \ref{HBs_ABchains_illuminations}). 
\begin{table}[h!]
\caption{\textsf{The 7 HBs between the optimized AB Chains or within its each Chain for each pdb model:}}
\centering
{\small
\begin{tabular}{l               |l                          |l}\hline
between B--A Chains             &in A Chain                 &in B Chain\\ \hline \hline
Asn171@H--Gln172@OE1            &                           &\\ 
                                &Asn174@OD1--Gln172@H       &Asn174@OD1--Gln172@H\\
                                &Asn173@H--Asn171@OD1       &Asn173@H--Asn171@OD1\\
                                &Asn170@H--Gln168@OE1       &Asn170@H--Gln168@OE1\\ \hline  
\end{tabular} 
} \label{HBs_ABchains}
\end{table}
The AB sheets stack in a parallel face-to-back configuration, the convex face of sheet A nestles against the concave back of its neighboring sheet B, with a high degree of surface complementarity (Fig. \ref{HBs_ABchains_illuminations}). Seeing Fig. \ref{6AXZ_ABchains_optimized_Hydrophobic_Polar}, in AB Chains, the most colored in pink are the polar residues Gln168--Asn174, another color of residues are the hydrophobic residues Phe175 and Val176. 

\begin{figure}[h!]
\centerline{
\includegraphics[width=5.6in]{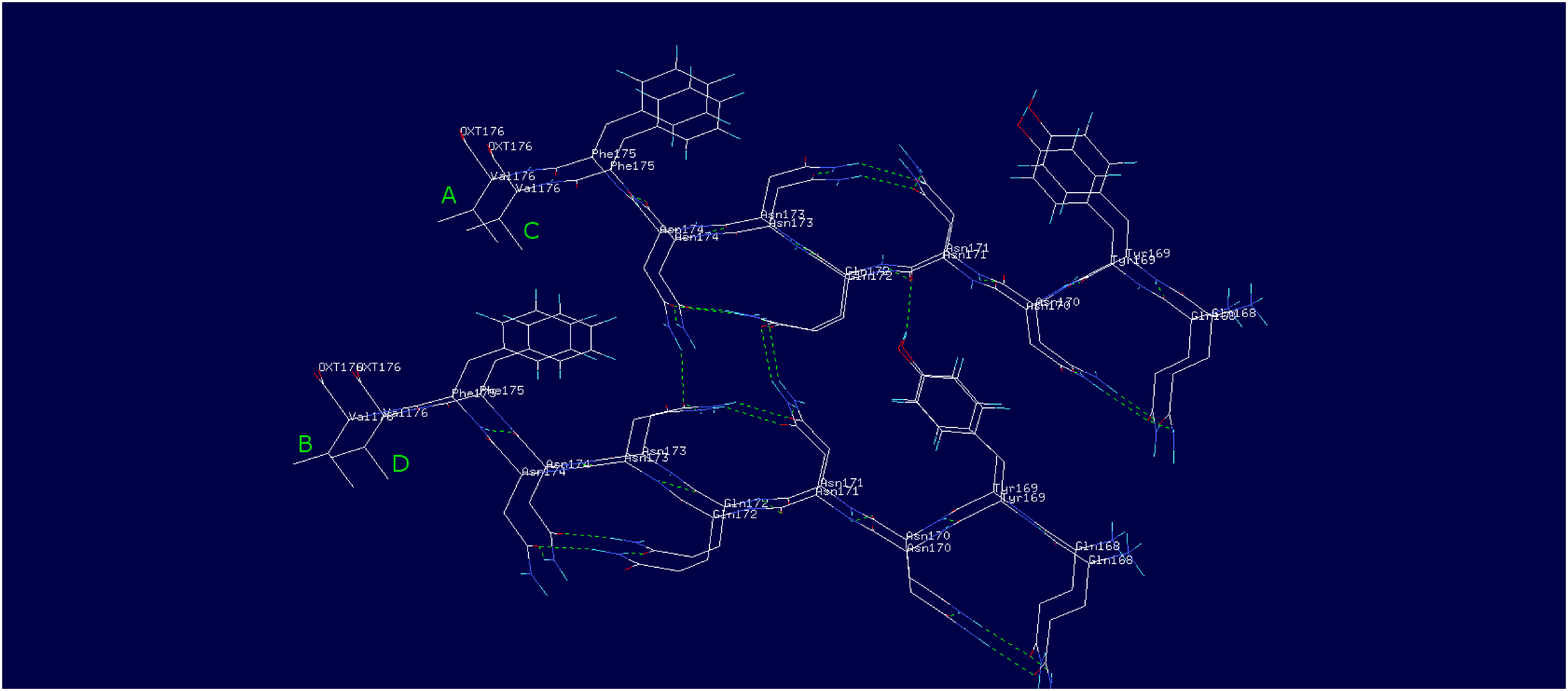}
}
\caption{\textsf{ The optimized ABCD Chains in the fibril structures of QYNNQNNFV segment 168--176 from bank vole prion (PDB entry 6AXZ). The dashed lines denote the HBs. The photos were produced by Swiss-PdbViewer 4.1.0.}} \label{HBs_ABCDchains_illuminations}
\end{figure}

{\it (III) ABCD-Chains}. In ABCD Chains, 28.9\% of the atoms are hydrogen. There are a lot of HBs for the optimized ABCD Chains of 6AXZ model (see Fig. \ref{HBs_ABCDchains_illuminations}, and the supplementary Tab. 2 of \cite{gallagher_etal2018} where B is the C in here and C is the B here), including the Asn171@H--Gln172@OE1 in Tab. \ref{HBs_ABchains} linking B and A Chains, the HBs Asn171@O--Tyr169@H and Asn174@H--Asn173@OD1 linking A and D Chains, and Gln172@OE1--Asn171@H linking C and D Chains (Tab. \ref{HBs_ABCDchains} and Fig. \ref{HBs_ABCDchains_illuminations}).
\begin{table}[h!]
\caption{\textsf{The 4 special HBs between the optimized ABCD Chains for the 6AXZ.pdb model:}}
\centering
{\small
\begin{tabular}{l               |l}\hline
The pair of Chains              &HBs\\ \hline \hline
BA             &Asn171@H--Gln172@OE1\\ \hline
AD             &Asn171@O--Tyr169@H, Asn174@H--Asn173@OD1\\ \hline
CD             &Gln172@OE1--Asn171@H\\ \hline
\end{tabular} 
} \label{HBs_ABCDchains}
\end{table}

\begin{figure}[h!]
\centerline{
\includegraphics[width=5.2in]{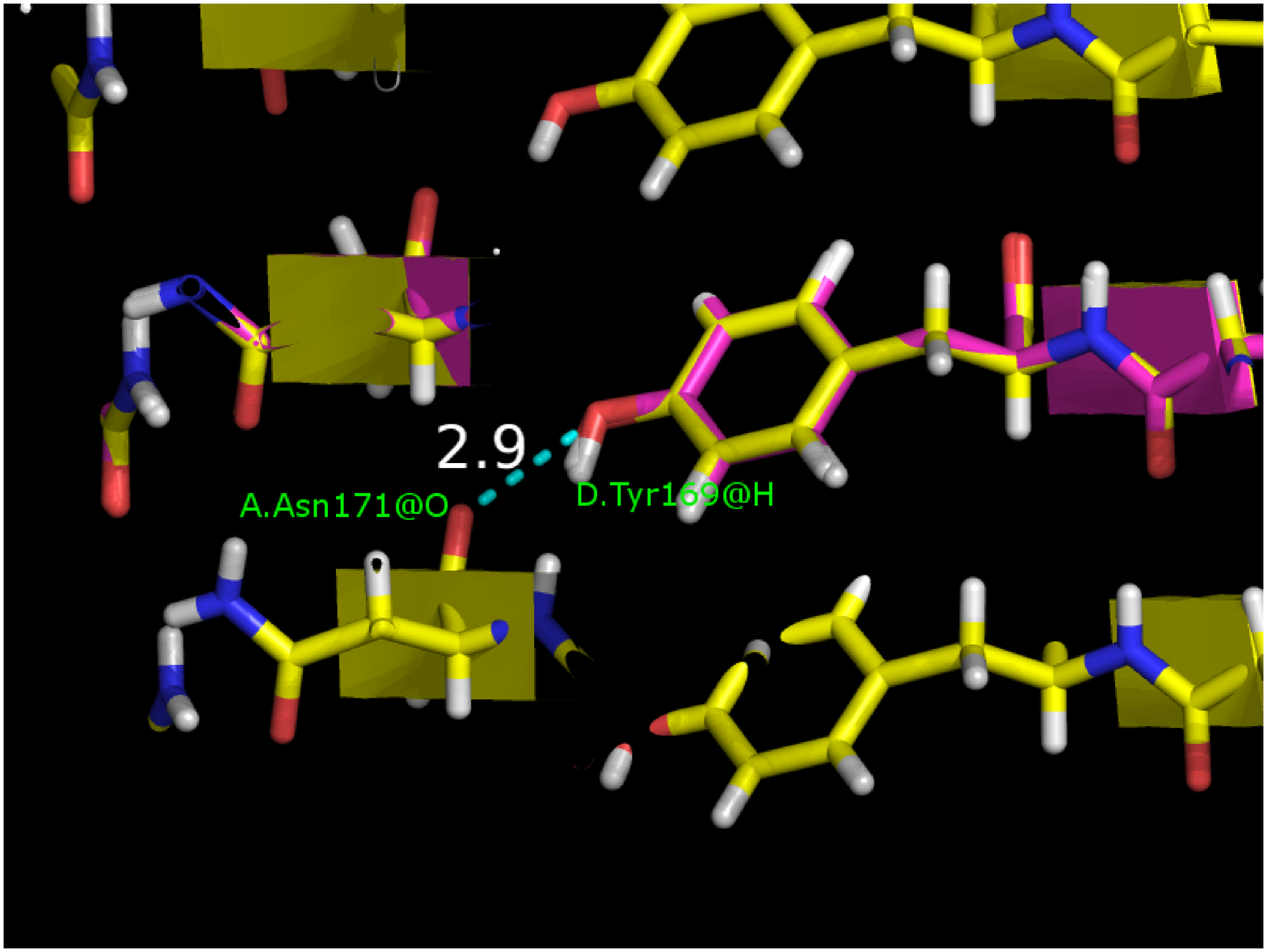}
}
\caption{\textsf{ The HB A.Asn171@O--D.Tyr169@H between the optimized AD Chains in the fibril structures of QYNNQNNFV segment 168--176 from bank vole prion (6AXZ.pdb). The dashed lines denote the HB. The photo was offered by Rodriguez JA (University of California Los Angeles).}} \label{HB_A_Asn171O-D_Tyr169H}
\end{figure}

By Figs. \ref{HBs_ABCDchains_illuminations}-\ref{HB_A_Asn171O-D_Tyr169H} and Tab. \ref{HBs_ABCDchains}, we can confirm the ``intersheet hydrogen bond between Tyr169 (of D Chain) and the backbone carbonyl of Asn171 (of A Chain) on an opposing strand" of \cite{gallagher_etal2018}, bearing in mind that
\begin{equation}
D = A + \left( \begin{array}{c}
-2.18510 + 4.94\\
10.10648\\ 
0\end{array} \right). \label{21_6AXZ_ad}
\end{equation}
 
At the end of this subsection, we revisit the HBs of the optimized ACE Chains. We illuminate them in Fig. \ref{HBs_ACEchains_illuminations}, where there are some `polar clasps' (as shown in the supplementary Fig. 7 of \cite{gallagher_etal2018}).  
\begin{figure}[h!]
\centerline{
\includegraphics[width=2.8in]{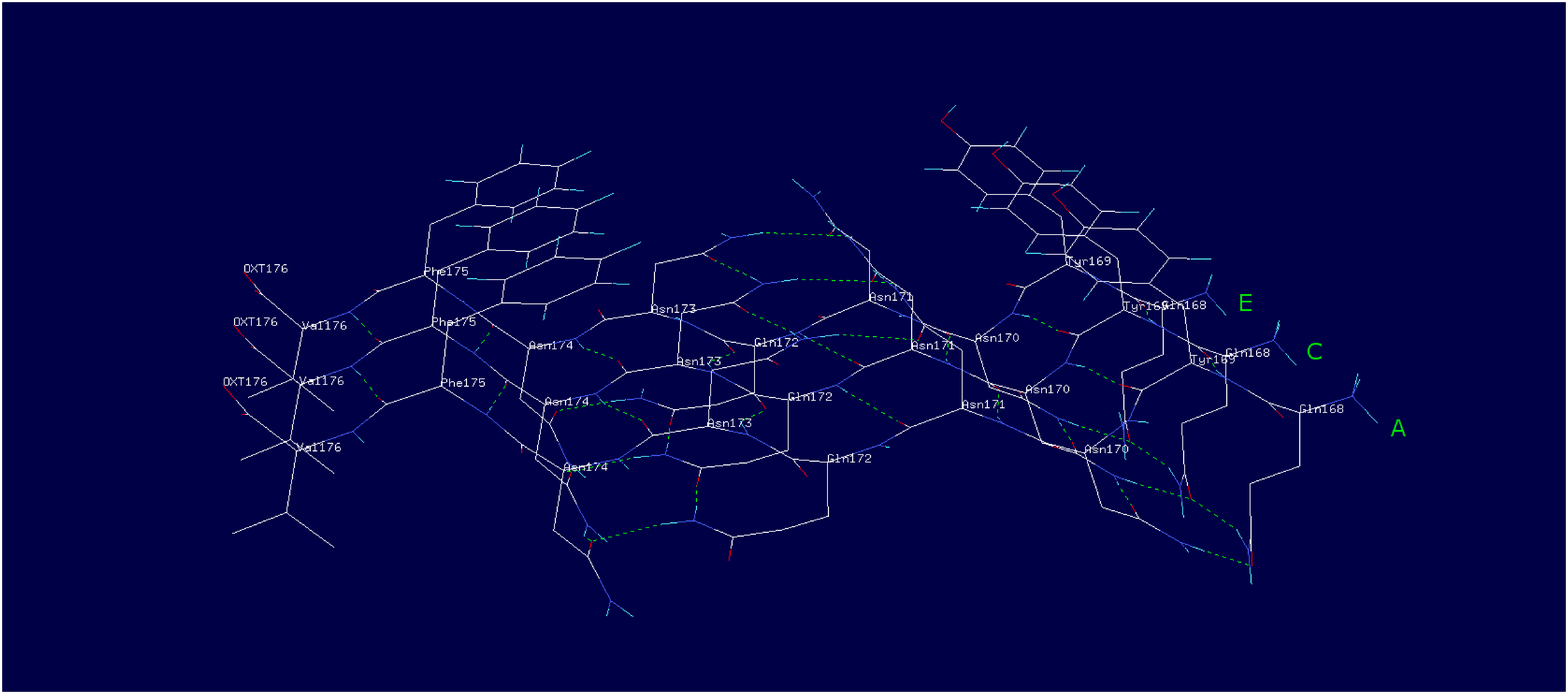}
\includegraphics[width=2.8in]{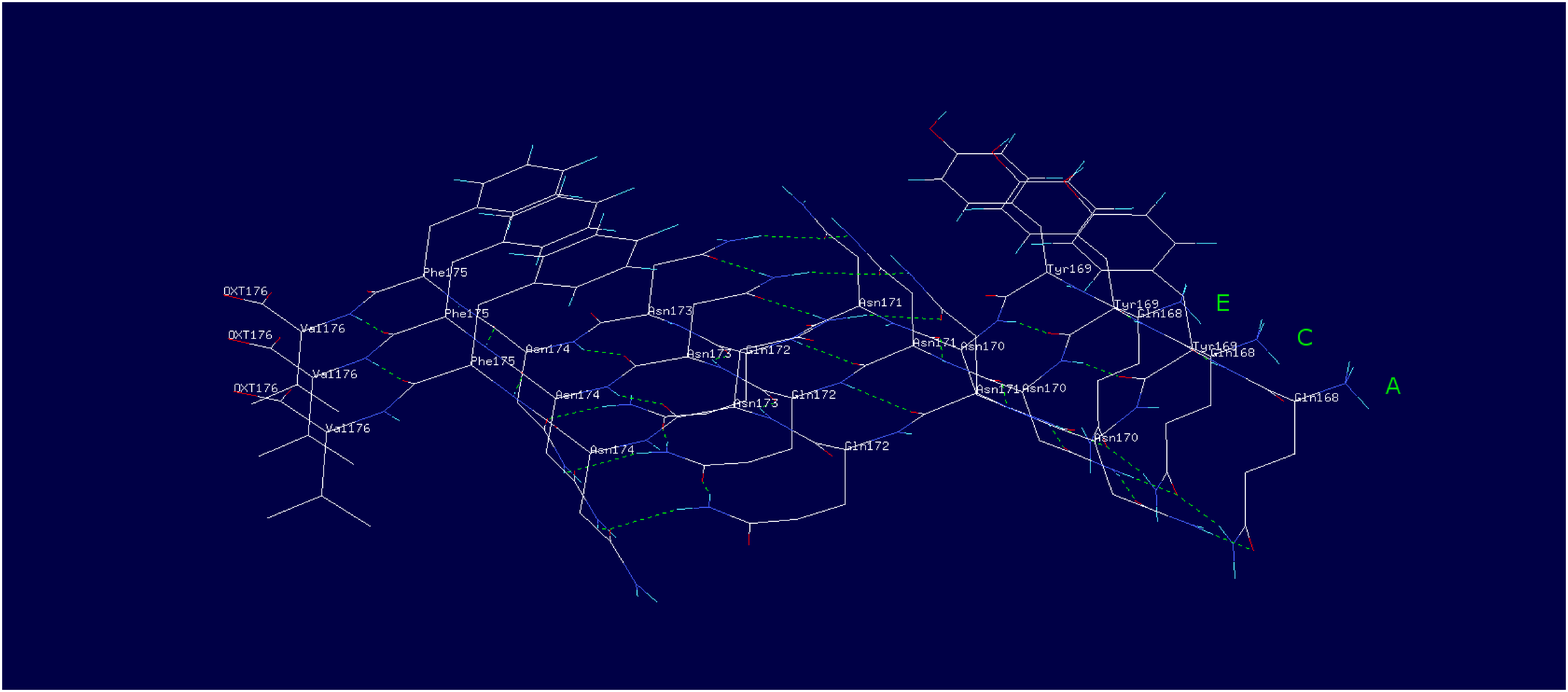}
}
\caption{\textsf{ The optimized ACE Chains in the fibril structures of QYNNQNNFV segment 168--176 from bank vole prion (left: 6AXZ, right: 6BTK). The dashed lines denote the HBs. The photos were produced by Swiss-PdbViewer 4.1.0. The green coloured A, C, E denote the A Chain, C Chain and E Chain respectively.}} \label{HBs_ACEchains_illuminations}
\end{figure}

\section{Summary}
This short paper revisited the `polar clasps' and presented detailed HBs of the `polar clasps'. We found out more HBs for the neighboring two $\beta$-strands, for the block of two $\beta$-sheets - each sheet being with two $\beta$-strands, and the most important, for the neighboring two single $\beta$-sheets AB, two new HBs B.Asn171@H--A.Gln172@OE1, B.Tyr169@OH--A.Gln172@N were found out, which is completely different from the experimental one (an intersheet HB between Tyr169 and the backbone carbonyl of Asn171 on an opposing strand) in \cite{gallagher_etal2018}.

The structures of some segments of PrP$^{Sc}$ have been determined, including the most recent bank vole PrP$^{Sc}$(168--176) one of \cite{gallagher_etal2018}. In Fig. 16.3 of \cite{zhang2015}, we have predicted that some segments PrP(126--133), PrP(137--143), PrP(168--176), PrP(170--175), PrP(177--182), PrP(211--216) should have the amyloid fibril forming property. 

This brief paper also showed us the importance of optimization. Before optimization, we could not find out any HBs (where the distance and angle cutoffs are the default values of VMD) from the structures calculated by mathematical formulas given according to crystal informatics in the PDB files. However, after optimization, all the HBs were revealed. The correct hybrid skill of optimization algorithms is also worth having a mention here. In the short comments, the author revisited two recently revealed Crystal structures of PrP segments with ``polar collapse" and found extra HBs in the network after optimization. The manuscript is clear written. Here we discuss more on comparing the structures in the PDB library and those after optimizations. How far away the current structures deposited in the PDB library from the refined structures are listed as follows. For 6AXZ.pdb, its RMSD \cite{zhang2018} values are 0.077589, 0.058435, 0.075521, and 0.098501 angstroms for ABCD Chains, AB Chains, AC Chains, and ACE Chains respectively. For 6BTK.pdb, its RMSD values are 0.056441, 0.045499, and 0.075953 angstroms for AB Chains, AC Chains, and ACE Chains respectively. These RMSD values could further highlight the importance/necessity of refinement in the protein structure study. 

In this paper, we revisited the `polar clasps' in the structures of PrP and presented detailed HBs of the ``polar clasps" after optimization. It is possible that different optimization strategies would have made different results. However, if the local searches in Swiss-PdbViewer 4.1.0 were carried out thoroughly, the optimization results should be the same ones.

\section*{Acknowledgments}
This research was supported by a Melbourne Bioinformatics grant numbered FED0001 on its Peak Computing Facility at the University of Melbourne, an initiative of the Victorian Government (Australia). The author thanks Rodriguez JA (University of California Los Angeles) for his helps to understand their paper \cite{gallagher_etal2018} deeply. The author is also grateful to some comments from reviewers, which have improved this paper greatly.

\end{document}